\documentclass[twocolumn,a4paper,superscriptaddress,prb]{revtex4}
\usepackage{amsmath}
\usepackage{graphicx}
\usepackage[english]{babel}

\begin{document}

\title{A comparative study of density functional and density functional tight binding calculations of defects in graphene}

\author{Alberto Zobelli}
\email{alberto.zobelli@u-psud.fr}
\affiliation{Laboratoire de Physique des Solides, Univ. Paris-Sud, CNRS-UMR
8502, 91405, Orsay, France}

\author{Viktoria Ivanovskaya}
\affiliation{Institut des Mat\'{e}riaux  Jean Rouxel (IMN), CNRS UMR 6502,
University of Nantes, 44322 Nantes, France}

\author{Philipp Wagner}
\affiliation{Institut des Mat\'{e}riaux  Jean Rouxel (IMN), CNRS UMR 6502,
University of Nantes, 44322 Nantes, France}

\author{Irene Suarez-Martinez}
\affiliation{Nanochemistry Research Institute, Curtin University of Technology,
Perth, Western Australia 6845, Australia}

\author{Abu Yaya}
\affiliation{Institut des Mat\'{e}riaux  Jean Rouxel (IMN), CNRS UMR 6502,
University of Nantes, 44322 Nantes, France}

\author{Chris P. Ewels}
\affiliation{Institut des Mat\'{e}riaux  Jean Rouxel (IMN), CNRS UMR 6502,
University of Nantes, 44322 Nantes, France}
\email{chris.ewels@cnrs-imn.fr}


\begin{abstract}
The density functional tight binding approach (DFTB) is well adapted for the
study of point and line defects in graphene based systems.  After briefly
reviewing the use of DFTB in this area, we present a comparative study of defect
structures, energies and dynamics between DFTB results obtained using the dftb+
code, and density functional results using the localised Gaussian orbital code,
AIMPRO.  DFTB accurately reproduces structures and energies for a range of point
defect structures such as vacancies and Stone-Wales defects in graphene, as well
as various unfunctionalised and hydroxylated graphene sheet edges.  Migration
barriers for the vacancy and Stone-Wales defect formation barriers are
accurately reproduced using a nudged elastic band approach.  Finally we explore
the potential for dynamic defect simulations using DFTB, taking as an example
electron irradiation damage in graphene.
\end{abstract}

\maketitle

\section{Introduction}

The family of carbon structures is much larger than its most notable
components and the number of new members synthesized each year makes it hard to
categorize all carbon forms.\cite{carbon-nanoforms}
Our ability to describe computationally the structure of realistic carbon systems faces the additional
difficulty represented by the presence of native defects, which often dominate the mechanical, electronic
and chemical properties of their host material. 
Furthermore, defect combinations can also serve as elemental
topological transformations that when applied to original perfect forms generate more complex structures. 

Modern electron microscopes can access subnanometric spatial
resolutions and thus are now able to image individual defects in nanostructures.
However image interpretation is not straightforward and accurate structural
models are still required for an in-depth understanding of defective atomic structures.
In this context computational simulations represent a necessary
complementary tool both for the interpretation of experimental data and for a deeper
understanding of the specific physical and chemical properties at defective sites.

Classical density functional theory (DFT) methods have been shown to describe
with a high accuracy the structure and electronics of defects in carbon
materials. Point defects can be simulated using either clusters or
periodic structures formed from hundred of atoms. However more extended
defects such as extended dislocation lines\cite{heggie},
turbostratic or misaligned graphite\cite{graphite-moire} or amorphous
carbon\cite{robertson1986amorphous} require significantly larger models. Until
recently
density functional techniques have been limited to carbon models containing only
few hundreds of atoms. This limitation is going to be overcome by very recent
improvements in filtration techniques on Gaussian basis sets that
can massively reduce computational time and memory
requirements.\cite{AIMPRO-filtration}
Thanks to these methodological developments combined with the continuous speeding up of
computing systems, routine study by full DFT of atomic structures with many thousands
of atoms is becoming a realistic task.

However, beside ground state determination, the study of the dynamics of defective
crystals remains too computationally intensive to tackle within the framework of current density
functional theory. Examples of these specific problems include the
growth mechanism of carbon nanostructures, their thermal or
mechanical stability, the effects of structural reorganization 
induced by high energy particle irradiation, and the diffusion of extended defects such as
vacancy clusters or dislocations. This range of problems are generally investigated using
molecular dynamics or Monte Carlo techniques where forces and energies are evaluated
using empirical or semi-empirical
approaches.\cite{irle2009milestones,petersen2005monte}
However many dynamical problems in carbon nanostructures involve mechanisms of bond
breaking and reconstruction that are usually poorly estimated by computational methods
parametrized on ground state configurations.

The usage of density functional tight binding (DFTB) for the study of dynamics
and reorganization of complex carbon structures represents an extremely helpful
compromise between accuracy and speed.\cite{Porezag-95,Seifert-96,Elstner-98}
Density functional tight binding parameters, in particular those
derived for carbon,\cite{Elstner-98} are highly transferable,
overcoming the main limitations of empirical and semi-empirical
techniques at the reduced computational cost of standard tight binding.
This is demonstrated by a wide range of problematics covered by a number of studies on
carbon based systems.
DFTB has been employed for instance in the study of the structure and energetics of point
defects in single walled carbon
nanotubes\cite{kotakoski2006energetics,sternberg2006carbon,malola2010structural,
krasheninnikov2004adsorption,cruz2010controlling,krasheninnikov2006bending}
and more extended defective structures such as screw dislocations in multi
walled carbon nanotubes\cite{irene-MWNTdislocations} and bonding between
fullerenes and nanocones.\cite{nanocones}
The capability of carbon DFTB parameters to reproduce complex rebydridization phenomena
has been shown for example in the simulation of the thermal induced
graphitization of nano-diamond surfaces,\cite{Diamwires} in monoatomic
carbon chains formation at axial strain applied to carbon
nanotubes,\cite{Vika-small} and
carbon nanostructure growth through C$_2$ addition.\cite{sacha}
DFTB molecular dynamics has been also employed in the study of
self-assembling processes of fullerene
cages\cite{doi:10.1021/nn900494s,irle2006c60,irle2003formation} and
nanotubes.\cite{ohta2008rapid}

In order to justify the use of DFTB approaches it is necessary to benchmark the results
against more conventional approaches such as full DFT calculations. Surprisingly
in the literature there is a lack of detailed comparative studies of this nature
for defective carbon nanomaterials and for
this reason we have undertaken the current work.
We have first chosen a range of well characterised
intrinsic point defect structures in graphene in order to benchmark the DFTB optimised geometries and formation
energies.  This is followed by a study of more complex unterminated and 
functionalised edges to quantify its capability with extended structural
defects. As well as static ground state structures it is important to analyse
the performance of the DFTB approach further from equilibrium and for this
reason we next study defect diffusion and formation/annihilation barriers. 
Finally we take as a complex dynamic example the formation of points defects in
carbon nanomaterials under the influence of an electron beam. As we show, DFTB
is capable of remarkably accurate reproduction of full DFT calculations
at a fraction of the computational cost, justifying its use in a wide range of structurally complex carbon nanoscale problems.

\section{Computational method}

DFT calculations are conducted using the AIMPRO2.0 code\cite{Aimpro,Aimp-2} and
the DFTB approximation as implemented in the dftb+
code\footnote{http://www.dftb-plus.info}, using comparable cells and k-point
meshes. 
 The AIMPRO calculations are performed under the local
density approximation using a localised Gaussian basis set with 22 independent
functions per carbon atom (12 per hydrogen and 40 per oxygen).  
Finite temperature Fermi smearing is used to control electron state population
near to the Fermi level with temperature kT=0.04~eV.  
Spin polarised calculation have been performed for open shell configurations
(carbon monovacancy, zigzag and Klein edges).
Pseudopotentials are taken from Hartwingser-Goedecker-Hutter.\cite{hgh}
DFTB parameters have been derived by M. Elstner et al.\cite{Elstner-98}
Both the DFT and DFTB calculations are fully self-consistent. 
No additional functionality such as Van der Waals corrections are used within
the DFTB calculations.

\section{Point and line defects in carbon nanostructures}

In this section we present a comparative study on the structure and formation energies of
topological defects in graphene and graphenic structures.  We take as our first test system a
series of standard intrinsic point defect structures in the graphene lattice, namely a single vacancy, 
a 5-8-5 divacancy pair, a ``Stone-Wales" defect (a rotation of two 
carbon atoms through 90$^\circ$ about their bond centre),
and an ``inverse Stone-Wales" defect (addition of a C$_2$ pair to the graphene
lattice across a single hexagon).  
Between them these defects contain a variety of local bonding including
undercoordinated carbon atoms,  dilated bonds, local out-of-plane distortions
and combinations of resonant and localised single-/double- bonds.  As such
they represent a stringent test for DFTB.
Through the use of an infinite graphene nanoribbon we then examine extended
defects in the  form of unterminated graphene sheet edges.  These once again
exhibit a range of bonding states including localised triple bond character
(armchair and reconstructed zigzag edges), extended metallic states (zigzag
edges) and singly coordinated carbon atoms (Klein edge).

\begin{figure*}
 \includegraphics[width=\textwidth]{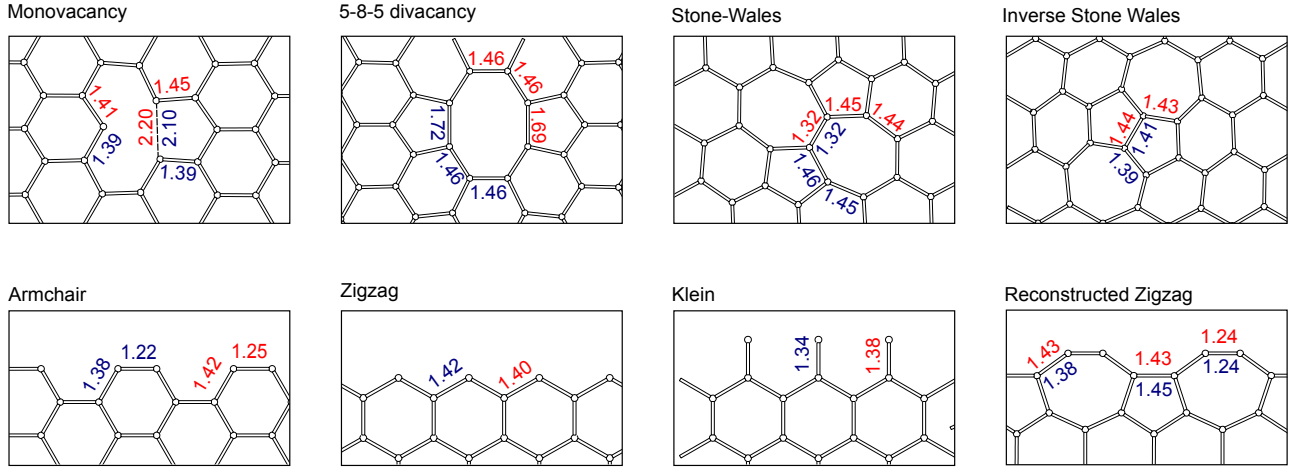}
 \caption{Structure of intrinsic point defects and unterminated edges in graphene. Values in the
figures represent bond lengths in \AA ngstroms obtained from DFT (blue) and DFTB (red).
DFT values from Reference.\cite{Vika-PRLedges}
}
\label{structures}
\end{figure*}

In figure \ref{structures} we present structural models for the different types of point
defects in graphene and several graphene edge configurations. In the figure we report
the most notable bond lengths as obtained after optimization with DFT (blue values) and
DFTB in its self consistent charge formulation (red values).  The DFT results
are detailed further in Ref.\cite{Vika-PRLedges}. 

The DFTB structures are in excellent agreement with the DFT data for all carbon coordinations,
from the single coordination of a carbon atom at a Klein edge, through the double coordinate
carbons at the zigzag and armchair edge or close to the vacancy site, to the modified triple
coordinated states as in the Stone-Wales and inverse Stone-Wales defect (C$_2$
addition). All DFTB-DFT bond lenght discrepancies are lower then 4\% and most
around
2\%.

Formation energies are presented in table,\ref{Point-def-energy} and once again
agreement is excellent,  with DFTB point defect formation energies matching DFT
values to within 1.5\% (and most less than 1\%).  The largest error is in the
Klein edge formation energy (19.7\%), which is understandable given that this is
a physically unstable edge structure (repetition of the unit cell and breakage
of the symmetry leads to spontaneous pairwise rebonding of the undercoordinated
atoms\cite{Vika-PRLedges}).  The other edge formation energies deviate from the
DFT values by 9.7\%, 1.8\% and 1.8\% respectively.

\begin{table}
\centering
 \begin{tabular}{lrr}
 \hline
 \hline
 \multicolumn{3}{c}{Point defects (eV)} \\
 \hline
& DFT & DFTB \\
 \hline
 Mono-vacancy & 7.40 & 7.51 \\
 5-8-5 Divacancy & 8.25 & 8.19 \\
 Stone-Wales & 4.86 & 4.85 \\
 Inverse Stone-Wales & 6.37 & 6.40\\
 \hline
 \hline
 \multicolumn{3}{c}{Edges (eV/\AA)} \\
 \hline
& DFT & DFTB \\
 \hline
 Zigzag edge & 1.34 & 1.21 \\
 Armchair edge & 1.10 & 1.08 \\
  Klein edge & 2.22 & 1.78\\
 Zigzag reconstructed edge & 1.09 & 1.07 \\
 -OH terminated armchair ribbon & -2.26 & -3.53\\
 \hline
 \hline
 \end{tabular}
 \caption{Formation energies for point defects and edges in graphene obtained by DFT and
DFTB.  DFT edge values from References \cite{Vika-PRLedges,Wagner2011}.}
 \label{Point-def-energy}
\end{table} 

Pristine graphene edges have dangling bonds at the edge atoms. Rehybridization
(as considered above) or H-termination is the simplest way to saturate these
dangling bonds, while inducing only small edge strain.
Edge functionalization by halogens or more complex functional
groups (-OH or -SH) has been shown to induce a significant strain along the
ribbon edge, through steric hindrance, electrostatic repulsion between groups,
inter-group bonding, etc.
Being energetically unfavourable, this strain can be
relieved via out-of-plane distortions. Specifically, hydroxyl (-OH) terminated
graphene nanoribbons of different widths have been shown to compensate the
induced strain by forming a localised out-of-plane static ripple along the
graphene sheet edge.\cite{Wagner2011} A key consequence of these functionalised
nanoribbon edges is that both
electronic and mechanical properties can be tuned.\cite{Wagner2011}

\begin{figure}
 \includegraphics[width=\columnwidth]{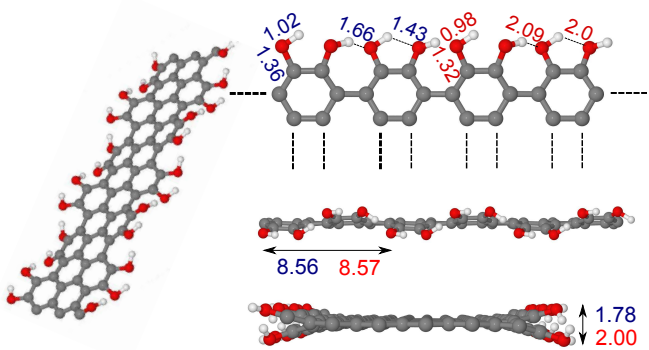}
 \caption{Structure of -OH terminated armchair graphene nanoribbon where values in
the figures represent bond length (in \AA) obtained from DFT (blue) and DFTB (red).
DFT values from Reference~\cite{Wagner2011}.}
 \label{figure_edge_ripple}
\end{figure}

In figure \ref{figure_edge_ripple} we present the structural model of an
hydroxyl
terminated graphene nanoribbon. Both DFT and DFTB calculations confirm that a
rippled configuration is more stable than any flat structure, allowing the
ribbon edge to relieve strain through out of plane
distortion.\cite{Wagner2011}

Once again, bond lengths obtained with DFT and DFTB generally correlate well.
The only discrepancy occurs at hydrogen bridges whose lengths are over-estimated
by DFTB.  This bond dilation is also reflected in the edge formation energy in
table \ref{Point-def-energy} where the DFTB edge formation energy is too
energetically stable compared to the DFT result.  DFTB is known to tend towards
overbinding for hydrogen-X bonds, and the dftb+ code includes a damping
correction to the short range contribution to the SCC interaction for
hydrogen\cite{gaus,yangelstner} which we did not use here.  In addition we can
also not
exclude the effect of the limited size of the DFTB basis or the possibility of
an incorrect estimation of the -OH group chemical potential, and further studies
are needed to fully explain this difference. 

In general we find a good correspondence in the energetics (formation energies)
and structural characteristics obtained by DFTB with DFT results, for both
intrinsic point and line defects in graphene, with reasonable structural and
energetic agreement for heteroatom systems given the limitations discussed
above.   These results support the use of DFTB in future, notably for problems
which can not be treated at the DFT level due to their size and complexity, or
due to the necessity for long trajectories, for example DFTB-MD
simulations on dynamics of graphene ribbon rippling (propagating and stationary waves along the edge). 

\section{Defect dynamics: formation and migration barriers}

The description of complex dynamical phenomena such as the reorganization of carbon
material when exposed to thermal treatments can be decomposed into several elementary
processes such as atom and vacancy migration and nucleation, bond rotation, and carbon atom
re-hybridization. Therefore, it is fundamental to derive accurately
minimum energy reaction paths and activation energies of these elemental
transformations in order to describe precisely more global transformations. However, the
ability of carbon to re-hybridize can render even elemental reaction paths highly complicated,
often introducing intermediate metastable configurations.

The nudged elastic band (NEB) method\cite{Mills-94,Henkelman-00} is a useful
techniques for
exploring in an efficient and automatic way a large region of the configuration space and
derive complex minimum energy paths. However, a large number of intermediate
images is generally required for obtaining with a good accuracy the saddle
point configuration and associated activation energies. Furthemore the NEB method
converges usually only after a large number of optimization steps. The high computational
cost of the NEB technique represents a strong limit for its usage in the framework of high
level computational approaches.

A good compromise can be obtained using the density functional tight binding theory
(DFTB). A combined NEB-DFTB approach has already been shown to give results
comparable to other higher-end techniques\cite{Fisher-05}. In the context of
layered
materials, DFTB-NEB
has been successfully employed in the study of single vacancies and vacancy complex
migration in boron nitride monolayers.\cite{Zobelli-migration}
Here we compare the possibilities of a DFTB-NEB approach with equivalent more
time consuming DFT-NEB calculations for the study of topological transformations in
carbon. Two example are presented where bond breaking occurs: mono-vacancy migration
and Stone-Wales bond rotation in a graphene plane.

\begin{table}
\centering
 \begin{tabular}{lcc}
 \hline
 \hline
 &\multicolumn{2}{c}{Barrier (eV)} \\ \hline
& DFT & DFTB \\
 \hline
 Vacancy Migration & 1.37 \cite{Diffusion-vacancies} & 1.29 \\
 Stone-Wales Formation & 9.2 \cite{defects-in-graphite}& 10.4 \\
 Stone-Wales Annihilation & 4.4 \cite{defects-in-graphite}& 4.7 \\
 \hline
 \hline
 \end{tabular}
 \caption{Calculated barriers (eV) for point defect formation and migration in graphene obtained by DFT and
DFTB using the nudged elastic band method.}
 \label{Diff-energy}
\end{table} 

Mono-vacancy diffusion in graphene occurs when a doubly coordinated carbon close to a
vacancy site breaks its two covalent bonds for rebonding with the opposite atom
pair neighboring the vacancy (the complete migration path also involves a
bonding rearrangement around the vacancy core but this has a very low activation
barrier). The minimum energy path saddle point obtained by DFTB-NEB
corresponds to a configuration where the migrating carbon atom lies in the middle point
between its initial and final position.  The graphene sheet undergoes a slight 
asymmetric out-of-plane deformation allowing the migrating atom to locate
itself
at the center of a compressed tetrahedron.
The DFTB activation energy we obtain is 1.29 eV, which is in an extremely good agreement
with the value of 1.37 eV obtained by an analogous DFT-NEB
study\cite{Diffusion-vacancies} (6\% underestimation). 

The 90 degree bond rotation required for the formation of a Stone-Wales defect
involves the breaking and reconstruction of two covalent bonds. Using DFTB-NEB
we obtain an activation energy for defect formation of 10.4 eV, with the
corresponding annihilation barrier for the reverse reaction of 4.7 eV.  These
compare extremely well to equivalent DFT values of 9.2 eV and 4.4 eV
respectively\cite{defects-in-graphite}, with DFTB thus overestimating the DFT
values by 13 and 7\% respectively.

These activation barrier calculations represent a stringent test for DFTB,
passing through structures which are far from the equilibrium defect ground
states. The excellent agreement between DFT and DFTB justifies the use of DFTB
in calculations of dynamic systems, and in the following section we show an
example of this where we use DFTB to examine atom loss during electron
irradiation of carbon nanostructures.

\section{Defect dynamics: electron irradiation in carbon nanostructures}

Electron irradiation is an unavoidable and generally unwanted side effect when high energy
electrons are used for imaging and analysis of nanostructures, such as in
transmission electron microscopy (TEM), but it can also be used to deliberately
restructure a carbon nanosystem in order to tune its mechanical and electronic
properties.
In this context it is desirable to be able to finely describe the
probability that a specific structural transformation occurs under electron
irradiation and how this probability depends on the energy of the electron beam.

Electron irradiation effects in carbon materials can be explained mostly by direct
elastic scattering between the relativistic electrons of the beam and the atomic
nucleus. For a given electron beam energy atoms can only be sputtered along
directions for which the transfered energy is above a certain emission
direction dependent energy threshold.

An analytic expression for the differential cross section as a function of the
emission direction has been derived by Mott\cite{Mott-32} and a useful
approximation of
the original expression has been obtained by McKinsley and
Feshbach.\cite{McKinley-48}
Total emission cross sections are derived by integrating the differential cross
section over the solid angle defined by the possible emission directions at a
given electron beam energy.

In Ref.\cite{Zobelli-irradiation} we proposed a methodology for deriving
anisotropic
emission energy threshold maps using extended molecular dynamics simulations. The
procedure consists in imparting an initial momentum to
the atom to sputter (direction and speed).  The system is successively allowed to evolve in a
microcanonical ensemble and at the end of the simulation the final position of the atom
is analyzed. The MD simulation is repeated increasing progressively the initial speed, up
to the critical limit for which the atom is ejected. The procedure is reiterated for a
number of different emission directions.

To obtain accurate energy threshold maps a high number of directions have to be
considered and the step size used to increase the initial velocity should be sufficiently small. In the
case of perfect graphene a map is obtained performing about 10000
molecular dynamics calculations, each of 200 MD steps on a structure containing
200 atoms. This number of calculations is too computationally demanding for
standard density functional techniques, but DFTB can produce a full emission map
at an affordable computational cost. This techniques has
thus be successfully used in the study of sputtering in perfect and
defective graphene and monolayer boron-nitride as well as in the study of irradiation
induced bond rotations in
graphene\cite{Zobelli-irradiation,Zobelli-shaping,kotakoski2010electron,
SW-kotakoski}.

\begin{figure}
\centering
\includegraphics[width=0.6\columnwidth]{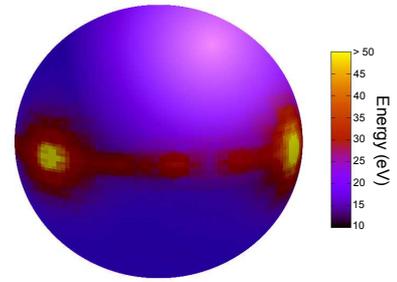}
\caption{Emission energy threshold map for a carbon atom in a graphene plane.
The spherical coordinate represent the emission direction and the color scale
the minimum emission kinetic energy. The equatorial plane corresponds to
emission directions on the graphene sheet, poles to directions orthogonal to the
sheet.}
\label{anysotropy}
\end{figure}

In figure \ref{anysotropy} we present the DFTB-MD derived emission energy
threshold map for a carbon atom from a
graphene plaine.
DFTB estimates the minimum ejection energy to be around 23 eV,
corresponding to an emission direction orthogonal to
the plane. A recent work has compared the DFT and DFTB emission energies 
obtaining, in the DFT case, a value of 22.2 eV.\cite{kotakoski2010electron} The
excellent agreement between the DFT and DFTB values (less than 4\% difference) validates the use of DFTB for
sputtering simulations in carbon materials.
However in the case of boron and nitrogen sputtering from a BN plane
the DFTB values are lower than the DFT. This discrepancy has been
discussed by the author as an inadequate description of charge transfer in DFTB
calculations for the BN system.\cite{kotakoski2010electron}

Considering the kinematic of the scattering problem carbon atoms can 
be sputtered from a graphene plane by electrons with an energy above 113 keV. 
Experimental TEM studies find an electron
beam energy limit situated between 90 and 100 keV. This discrepancy can be reasonably
attributed to the well known dissociation energy overestimation occurring in DFT-LDA that
also affects the DFTB parameters.
This error can be considered as systematic and the theoretical results be can corrected by
recalibrating on the experimental values.

\begin{figure}
\centering
\includegraphics[width=0.8\columnwidth]{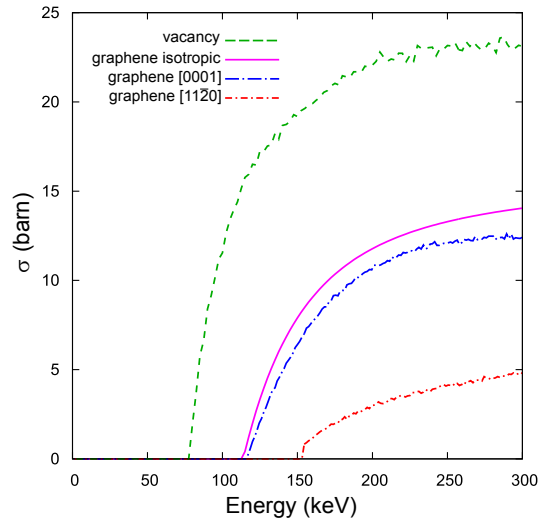}
 \caption{Sputtering cross sections for a carbon atom in a graphene plane using
the isotropic emission threshold approximation and the anisotropic approximation
when the electron beam direction is orthogonal or parallel to the plane.
Sputtering cross sections
for a carbon atom neighbouring a pre-existing vacancy in graphene (beam
direction orthogonal)}
\label{crossections}
\end{figure}

In Fig. \ref{crossections} we present the total knock-on cross section as a function of
the electron beam energy for a carbon atom in a graphene plane. 
A commonly used approximation considers the emission energy threshold as
independent of the emission direction, an analytic expression
 for the total cross section can then be found.\cite{Banhart-99}
This assumption brings however, as shown in Fig. \ref{crossections}, to
overestimate the integration region and thus the emission cross sections and it
cannot take into account the strong variation of cross section as a function of
the electron beam orientation in respect to the graphene.
In figure \ref{crossections} we present also the cross section for a carbon
atom neighboring a pre-existing vacancy site. The reduced coordination of the
knocked carbon atom makes sputtering more probable than for an atom in perfect
graphene. This higher cross section explains the vacancy clustering in graphene
observed by TEM.

Calculated cross sections have been used to optimize irradiation conditions
in a scanning transmission electron microscope for reshape individual single
walled carbon nanotubes at a nanometrical scale.\cite{Zobelli-shaping}
Cross section values can also be used in kinetic Monte Carlo simulations
of the global transformation of carbon structures under electron irradiation:
calculated cross sections attribute a sputtering probability
to each carbon atom, DFTB can successively be employed to optimize the system
after each ejection event.

\section{Conclusions}

The versatility of carbon in its range of bonding leads to a rich variety of low symmetry
materials, structures and defects.  However the corresponding size of the
resultant calculations, the range of minima to explore,
and the complex energy surfaces in non-equilibrium situations render many such problems outside the 
scope of conventional density functional approaches.
We show in the current study that the density functional tight binding approach
is able to successfully reproduce, with high quantitative accuracy, both
structural and energetic data from full density functional calculations at a
fraction of the computational cost.  Calculations of atom knock-on cross
sections under electron irradiation provide an example where DFTB calculations
are able to advance and guide our manipulation of carbon materials at the atomic
scale. These results confirm that DFTB represents a powerful tool for
computationally intensive studies of carbon nanomaterials. 

\acknowledgements

We would like to thank R. Barthel for the implementation
of the NEB technique using DFTB, and B. Hourahine for useful discussions.  CPE, VI and PW acknowledge French ANR
P3N project ANR-09-NANO-016-04 ``Nanosim\_Graphene" for funding.  All authors thank COST
project MP0901 ``NanoTP" for support.

\bibliographystyle{apsrev}
\bibliography{AZ-pssb}

\providecommand{\WileyBibTextsc}{}
\let\textsc\WileyBibTextsc
\providecommand{\othercit}{}
\providecommand{\jr}[1]{#1}
\providecommand{\etal}{~et~al.}


\begin{thebibliography}{[10]}

\bibitem{carbon-nanoforms}
 \textsc{I.~Suarez-Martinez},  \textsc{N.~Grobert},  and
  \textsc{C.~Ewels},
 \jr{Carbon} p.\,In Press (2011),
doi:10.1016/j.carbon.2011.11.002.


\bibitem{heggie}
 \textsc{R.\,H. Telling} and  \textsc{M.\,I. Heggie},
 \jr{Philosophical Magazine Letters} \textbf{83}(7), 411 (2003).


\bibitem{graphite-moire}
 \textsc{J.\,M. Campanera},  \textsc{G.~Savini},  \textsc{I.~Suarez-Martinez},
  and  \textsc{M.\,I. Heggie},
 \jr{Phys. Rev. B} \textbf{75}(Jun), 235449 (2007).


\bibitem{robertson1986amorphous}
 \textsc{J.~Robertson},
 \jr{Advances in Physics} \textbf{35}(4), 317 (1986).


\bibitem{AIMPRO-filtration}
 \textsc{M.\,J. Rayson} and  \textsc{P.\,R. Briddon},
 \jr{Phys. Rev. B} \textbf{80}, 205104 (2009).


\bibitem{irle2009milestones}
 \textsc{S.~Irle},  \textsc{Y.~Ohta},  \textsc{Y.~Okamoto},  \textsc{A.~Page},
  \textsc{Y.~Wang},  and  \textsc{K.~Morokuma},
 \jr{Nano Research} \textbf{2}(10), 755--767 (2009).


\bibitem{petersen2005monte}
 \textsc{T.~Petersen},  \textsc{I.~Snook},  \textsc{I.~Yarovsky},  and
  \textsc{D.~McCulloch},
 \jr{Physical Review B} \textbf{72}(12), 125417 (2005).


\bibitem{Porezag-95}
 \textsc{D.~Porezag},  \textsc{T.~Frauenheim},  \textsc{T.~K\"ohler},
  \textsc{G.~Seifert},  and  \textsc{R.~Kaschner},
 \jr{Phys. Rev. B} \textbf{51}, 12947 (1995).


\bibitem{Seifert-96}
 \textsc{G.~Seifert},  \textsc{D.~Porezag},  and  \textsc{T.~Frauenheim},
 \jr{Int. J. Quantum Chem.} \textbf{58}, 185 (1996).


\bibitem{Elstner-98}
 \textsc{M.~Elstner},  \textsc{D.~Porezag},  \textsc{G.~Jungnickel},
  \textsc{J.~Elsner},  \textsc{M.~Haugk},  \textsc{T.~Frauenheim},
  \textsc{S.~Suhai},  and  \textsc{G.~Seifert},
 \jr{Phys. Rev. B} \textbf{58}, 7260 (1998).


\bibitem{kotakoski2006energetics}
 \textsc{J.~Kotakoski},  \textsc{A.~Krasheninnikov},  and
  \textsc{K.~Nordlund},
 \jr{Phys. Rev. B} \textbf{74}(24), 245420 (2006).


\bibitem{sternberg2006carbon}
 \textsc{M.~Sternberg},  \textsc{L.~Curtiss},  \textsc{D.~Gruen},
  \textsc{G.~Kedziora},  \textsc{D.~Horner},  \textsc{P.~Redfern},  and
  \textsc{P.~Zapol},
 \jr{Phys. Rev. Lett.} \textbf{96}(7), 75506 (2006).


\bibitem{malola2010structural}
 \textsc{S.~Malola},  \textsc{H.~H{\"a}kkinen},  and
  \textsc{P.~Koskinen},
 \jr{Phys. Rev. B} \textbf{81}(16), 165447 (2010).


\bibitem{krasheninnikov2004adsorption}
 \textsc{A.\,V. Krasheninnikov},  \textsc{K.~Nordlund},  \textsc{P.\,O.
  Lehtinen},  \textsc{A.\,S. Foster},  \textsc{A.~Ayuela},  and  \textsc{R.\,M.
  Nieminen},
 \jr{Carbon} \textbf{42}(5-6), 1021 (2004).


\bibitem{cruz2010controlling}
 \textsc{E.~Cruz-Silva},  \textsc{A.~Botello-M{\'e}ndez},  \textsc{Z.~Barnett},
   \textsc{X.~Jia},  \textsc{M.~Dresselhaus},  \textsc{H.~Terrones},
  \textsc{M.~Terrones},  \textsc{B.~Sumpter},  and  \textsc{V.~Meunier},
 \jr{Phys. Rev. Lett.} \textbf{105}(4), 45501 (2010).


\bibitem{krasheninnikov2006bending}
 \textsc{A.\,V. Krasheninnikov},  \textsc{P.\,O. Lehtinen},  \textsc{A.\,S.
  Foster},  and  \textsc{R.\,M. Nieminen},
 \jr{Chem. Phys. Lett.} \textbf{418}(1-3), 132 (2006).


\bibitem{irene-MWNTdislocations}
 \textsc{I.~Suarez-Martinez},  \textsc{G.~Savini},  \textsc{A.~Zobelli},  and
  \textsc{M.~Heggie},
 \jr{J. Nanosci. and Nanotechnol.} \textbf{7}(10), 3417 (2007).


\bibitem{nanocones}
 \textsc{I.~Suarez-Martinez},  \textsc{M.~Monthioux},  and  \textsc{C.\,P.
  Ewels},
 \jr{J. Nanosci. Nanotech.} \textbf{9}, 6144 (2009).


\bibitem{Diamwires}
 \textsc{V.~Ivanovskaya} and  \textsc{A.~Ivanovskii},
 \jr{Inorganic Materials} \textbf{43}, 349 (2007).


\bibitem{Vika-small}
 \textsc{V.~Ivanovskaya},  \textsc{N.~Ranjan},  \textsc{T.~Heine},
  \textsc{G.~Merino},  and  \textsc{G.~Seifert},
 \jr{Small} \textbf{1}(4), 399 (2005).


\bibitem{sacha}
 \textsc{C.\,P. Ewels},  \textsc{G.~Van~Lier},  \textsc{P.~Geerlings},  and
  \textsc{J.\,C. Charlier},
 \jr{J. Chem. Inf. Model.} \textbf{47}(6), 2208 (2007).


\bibitem{doi:10.1021/nn900494s}
 \textsc{B.~Saha},  \textsc{S.~Shindo},  \textsc{S.~Irle},  and
  \textsc{K.~Morokuma},
 \jr{ACS Nano} \textbf{3}(8), 2241 (2009).


\bibitem{irle2006c60}
 \textsc{S.~Irle},  \textsc{G.~Zheng},  \textsc{Z.~Wang},  and
  \textsc{K.~Morokuma},
 \jr{The Journal of Physical Chemistry B} \textbf{110}(30), 14531 (2006).


\bibitem{irle2003formation}
 \textsc{S.~Irle},  \textsc{G.~Zheng},  \textsc{M.~Elstner},  and
  \textsc{K.~Morokuma},
 \jr{Nano Letters} \textbf{3}(4), 465 (2003).


\bibitem{ohta2008rapid}
 \textsc{Y.~Ohta},  \textsc{Y.~Okamoto},  \textsc{S.~Irle},  and
  \textsc{K.~Morokuma},
 \jr{ACS Nano} \textbf{2}(7), 1437 (2008).


\bibitem{Aimpro}
 \textsc{R.~Jones} and  \textsc{P.~Briddon},
 \jr{Semicond. Semimetals} \textbf{51A}, 287 (1998).


\bibitem{Aimp-2}
 \textsc{M.~Rayson} and  \textsc{P.~Briddon},
 \jr{Comp. Phys. Comm.} \textbf{178}, 128 (2008).


\bibitem{hgh}
 \textsc{C.~Hartwigsen},  \textsc{S.~Goedecker},  and
  \textsc{J.~H\"utter},
 \jr{Phys. Rev. B.} \textbf{58}, 3641 (1998).


\bibitem{Vika-PRLedges}
 \textsc{V.\,V. Ivanovskaya},  \textsc{A.~Zobelli},  \textsc{P.~Wagner},
  \textsc{M.\,I. Heggie},  \textsc{P.\,R. Briddon},  \textsc{M.\,J. Rayson},
  and  \textsc{C.\,P. Ewels},
 \jr{Phys. Rev. Lett.} \textbf{107}, 065502 (2011).


\bibitem{Wagner2011}
 \textsc{P.~Wagner},  \textsc{C.\,P. Ewels},  \textsc{V.\,V. Ivanovskaya},
  \textsc{P.\,R. Briddon},  \textsc{A.~Pateau},  and
  \textsc{B.~Humbert},
 \jr{Phys. Rev. B} \textbf{84}, 134110 (2011).


\bibitem{gaus}
 \textsc{M.~Gaus},  \textsc{M.~Cui},  and  \textsc{J.~Elstner},
 \jr{J. Chem. Theory Comput.} \textbf{7}, 931--948 (2011).


\bibitem{yangelstner}
 \textsc{Y.~Yang},  \textsc{H.~Yu},  \textsc{D.~York},  \textsc{Q.~Cui},  and
  \textsc{J.~Elstner},
 \jr{J. Phys. Chem. A} \textbf{111}, 10861 (2007).


\bibitem{Mills-94}
 \textsc{G.~Mills} and  \textsc{H.~J\'onsson},
 \jr{Phys. Rev. Lett.} \textbf{72}, 1124 (1994).


\bibitem{Henkelman-00}
 \textsc{G.~Henkelman},  \textsc{B.~Uberuaga},  and  \textsc{H.~J\'onnson},
 \jr{J. Chem. Phys.} \textbf{113}, 9901 (2000).


\bibitem{Fisher-05}
 \textsc{G.~Fischer},  \textsc{R.~Barthel},  and  \textsc{G.~Seifert},
 \jr{Eur. Phys. J. D} \textbf{35}, 479 (2005).


\bibitem{Zobelli-migration}
 \textsc{A.~Zobelli},  \textsc{C.\,P. Ewels},  \textsc{A.~Gloter},  and
  \textsc{G.~Seifert},
 \jr{Phys. Rev. B} \textbf{75}, 094104 (2007).


\bibitem{Diffusion-vacancies}
 \textsc{H.~Zhang},  \textsc{M.~Zhao},  \textsc{X.~Yang},  \textsc{H.~Xia},
  \textsc{X.~Liu},  and  \textsc{Y.~Xia},
 \jr{{Diam. Relat. Mater.}} \textbf{{19}}({10}), {1240--1244} ({2010}).


\bibitem{defects-in-graphite}
 \textsc{L.~Li},  \textsc{S.~Reich},  and  \textsc{J.~Robertson},
 \jr{Phys. Rev. B} \textbf{72}, 184109 (2005).


\bibitem{Mott-32}
 \textsc{N.~Mott},
 \jr{Proc. Roy. Soc. A} \textbf{135}, 429 (1932).


\bibitem{McKinley-48}
 \textsc{W.~McKinsley} and  \textsc{H.~Feshbach},
 \jr{Phys. Rev.} \textbf{74}, 1759 (1948).


\bibitem{Zobelli-irradiation}
 \textsc{A.~Zobelli},  \textsc{A.~Gloter},  \textsc{C.\,P. Ewels},
  \textsc{G.~Seifert},  and  \textsc{C.~Colliex},
 \jr{Phys. Rev. B} \textbf{75}, 245402 (2007).


\bibitem{Zobelli-shaping}
 \textsc{A.~Zobelli},  \textsc{A.~Gloter},  \textsc{C.\,P. Ewels},  and
  \textsc{C.~Colliex},
 \jr{Phys. Rev. B} \textbf{77}, 045410 (2008).


\bibitem{kotakoski2010electron}
 \textsc{J.~Kotakoski},  \textsc{C.~Jin},  \textsc{O.~Lehtinen},
  \textsc{K.~Suenaga},  and  \textsc{A.~Krasheninnikov},
 \jr{Physical Review B} \textbf{82}(11), 113404 (2010).


\bibitem{SW-kotakoski}
 \textsc{J.~Kotakoski},  \textsc{J.\,C. Meyer},  \textsc{S.~Kurasch},
  \textsc{D.~Santos-Cottin},  \textsc{U.~Kaiser},  and  \textsc{A.\,V.
  Krasheninnikov},
 \jr{Phys. Rev. B} \textbf{83}, 245420 (2011).


\bibitem{Banhart-99}
 \textsc{F.~Banhart},
 \jr{Rep. Prog. Phys.} \textbf{62}, 1181 (1999).


\end{thebibliography}

\end{document}